\documentclass[prl,floatfix,twocolumn,showpacs,superscriptaddress]{revtex4-1}
\usepackage{color,soul}
\usepackage{graphicx}
\usepackage{amsmath, amsthm, amssymb}
\usepackage{amsthm}
\usepackage{multirow}
\usepackage[normalem]{ulem}
\usepackage{soul}
\newcommand{\be}{\begin{equation}}
\newcommand{\ee}{\end{equation}}
\newcommand{\bea}{\begin{eqnarray}}
\newcommand{\eea}{\end{eqnarray}}
\begin{document}
\title{
Mechanism of chain collapse of strongly charged polyelectrolytes
}
\author{Anvy Moly Tom}
\email{anvym@imsc.res.in}
\affiliation{The Institute of Mathematical Sciences, C.I.T. Campus,
Taramani, Chennai 600113, India}
\author{Satyavani Vemparala}
\email{vani@imsc.res.in}
\affiliation{The Institute of Mathematical Sciences, C.I.T. Campus,
Taramani, Chennai 600113, India}
\author{R. Rajesh}
\email{rrajesh@imsc.res.in}
\affiliation{The Institute of Mathematical Sciences, C.I.T. Campus,
Taramani, Chennai 600113, India}
\author{Nikolai V. Brilliantov}
\email{nb144@leicester.ac.uk}
\affiliation{Department of Mathematics, University of Leicester, Leicester LE1 7RH,
United Kingdom}

\date{\today}

\begin{abstract}
We perform extensive molecular dynamics simulations of a charged polymer in a good solvent in the regime where the chain is collapsed. We analyze the dependence of the gyration radius $R_g $ on the reduced Bjerrum length $\ell_B$ and find two different regimes. In the first one, called as a weak electrostatic regime, $R_g \sim \ell_B^{-1/2}$, which is consistent only with the predictions of the counterion-fluctuation theory. In the second one, called a strong electrostatic regime, we find $R_g \sim \ell_B^{-1/5}$. To explain the novel regime we modify the counterion-fluctuation theory.
\end{abstract}
\maketitle

\emph{Introduction.} The  conformational states of a flexible neutral polymer in different solvents are well known.  It is extended  in a good solvent due to favorable excluded volume
interactions with the solvent molecules and collapses into a compact globule  in a bad solvent~\cite{FloryBook,Netz2003,khokhlov1994}. In contrast, a flexible polyelectrolyte (PE) -- charged polymer in the presence of counterions -- undergoes an extended to collapsed transition in both good and  bad solvents. Unlike neutral polymers, the conformations of a PE depend not only on the solvent quality, but also crucially on the interplay between electrostatic energy and
translational entropy of counterions~\cite{brilliantov98,Dobrynin2005}.
The strength of the electrostatic interactions depends on the charge density along the PE which
is quantified by the dimensionless Bjerrum length $\ell_B$.
For small charge density, counterions are dispersed away from the PE,  and the chain is in an extended necklace conformation when in a good or theta-solvent~\cite{khokhlov1994} and is collapsed into a compact globule in a bad solvent~\cite{anoop11,khokhlov1994}.  With increasing charge density, the PE attains an extended conformation, regardless of the solvent quality and
counterions begin
to condense onto the PE, renormalizing its charge density~\cite{manning69,Khokhlov1980,khokhlov1994}.
Further increase of the PE charge density results in an effective
attraction between similarly-charged monomers of the PE and  it collapses into a globule conformation, independent of the
solvent quality~\cite{Kremer1993,Kremer1995,winkler98,brilliantov98,anoop11,Dobrynin2005,Chertovich2016}.

The compaction of a PE chain into a globular conformation is of  great biological importance.  For instance, biological PEs like RNA or DNA are densely packed in cells and
viruses~\cite{VirusRNA2006,VirusElectro2012,VirusGrossberg2016}
which are orders of magnitude smaller than the contour length of the PE, requiring it to be highly compacted~\cite{Wong-Rev,delaCruz2000}.
The understanding of  DNA compaction is thus crucial for future gene therapy and production of synthetic cells.
Furthermore, the effective interactions driving the collapse of a single PE chain are closely connected to those resulting in aggregation of rigid PEs~\cite{anoop12,anvy16}.
Common
biological polymers like DNA, actin and microtubules are examples of rigid PEs whose aggregates play an important role in functions like cell scaffolding making it vital to understand the nature of attractive forces between  similar charges~\cite{Levin2002}.

To describe the counterintuitive phenomenon of  PE collapse in a good solvent, several
competing theories have been proposed~\cite{brilliantov98,pincus98,Kardar1999,delaCruz2000,Cherstvy2010,muthukumar2004,arti14}. 
The first theory is based on modelling  the collapsed conformation  as an amorphous ionic solid~\cite{delaCruz2000}. For large charge density of the PE and in the presence of
multivalent counterions the free energy of the solid is smaller than that of the extended PE, driving the chain collapse.
This theory, however, does not predict any dependence of the gyration radius $R_g$ on $\ell_B$.
In the second  group of theories,
it is assumed that condensed counterions and the PE monomers form dipoles~\cite{pincus98,Cherstvy2010,muthukumar2004,arti14}. The dipoles freely rotate yielding,  on average, an attractive interaction between the segments of the chain; this leads  to collapse of a PE even in a good solvent. For a highly  charged flexible PE in a salt-free solution, this theory predicts that the  radius of  gyration
of the collapsed conformation  scales as  ${R_g} \sim  N^{1/3}\left| \ell_B^2 - c B \right|^{-1/3} $, where $B$ is the second virial coefficient, $N$ is the number of chain monomers and $c$ is a dimensional constant that depends on the details of the system~\footnote{$c$ is different in different theories of Refs.~\cite{pincus98,muthukumar2004,arti14}}. This dependence is predicted for both good~\cite{pincus98,muthukumar2004,arti14} and bad~\cite{pincus98,arti14} solvents. For theta-solvent with $B=0$~\cite{khokhlov1994}, a  simpler scaling, $R_g \sim  {\ell_B^{-2/3} N{^{1/3}}}$  is obtained~\cite{pincus98}.

Finally, the third theory, referred to as counterion-fluctuation theory,  argues  that the collapse of a PE is due to negative pressure arising from fluctuations in the density of condensed counterions, which move freely  within a  PE globule~\cite{brilliantov98}. Such a physical picture of the condensed counterion motion agrees with the recent results of molecular dynamics (MD) simulations~\cite{Chertovich2016}. The counterion-fluctuation theory, when restricted to the second virial coefficient, predicts  that in a good solvent $R_g \sim  {\ell_B^{-1/2} N{^{1/3}}}$. Note that all mechanisms discussed above imply a collapsed phase, $R_g \sim N^{1/3}$ for large charge density, but different dependence of $R_g$ on $\ell_B$.

Due to a great significance for applications, especially for nano-medicine and biotechnology, it is vital to have an appropriate theory of the interactions that drive the collapse of a PE. In this Letter, we report the results of extensive MD simulations exploring the collapsed conformation of a single flexible PE chain in a good solvent. Two regimes in the dependence $R_g(\ell_B)$ have been numerically revealed: The one, consistent with the counterion fluctuation theory, $R_g \sim  {\ell_B^{-1/2}}$ \cite{brilliantov98},  and the new regime,
$R_g \sim  {\ell_B^{-1/5}}$, which we explain modifying the above theory.

\emph{MD simulations.} We model a flexible PE chain as $N$ monomers of charge $e$ ($e>0$ is the elementary charge) connected by harmonic springs of energy,
\be
U_{bond}(r)=\frac{1}{2} k(r-a)^2,
\ee
where $k$ is the spring constant, $a$ is the equilibrium bond length and $r$ is the distance between the bonded monomers. The chain and $N_c=N/Z$ neutralizing counterions, each of charge $-Z e$, with $Z=1,2,3$  being  the valency, are placed in a box of linear size $L$.
Pairs of all non-bonded particles  (counterions and monomers) separated by a distance $r$ interact through the $6-12$ Lennard Jones potential cutoff at $r_c$:
\begin{equation}
U_{LJ}(r)=4\epsilon \left[\left(\sigma/r \right)^{12}-\left(\sigma/r \right)^6 \right].
\end{equation}
The values of $\epsilon$ and $r_c$ are varied depending on the system being simulated. The electrostatic energy between charges $q_i$ and $q_j$ separated by $r_{ij}$ is
\begin{equation}
U_{c}(r_{ij})= \frac{q_iq_j}{\varepsilon r_{ij}},
\label{eq.1}
\end{equation}
where $\varepsilon$ is the dielectric permittivity of the solution.
The charge density along the PE chain is parameterized by the
dimensionless Bjerrum length $\ell_B$~\cite{khokhlov1994}:
\begin{equation}
\ell_B=\frac{1}{a}\frac{e^{2}}{(\varepsilon k_{B}T)}=\frac{\beta e^2}{\varepsilon a},
\label{eq.4}
\end{equation}
where $k_B$ is  the Boltzmann constant, $T$ is temperature and $\beta=(k_BT)^{-1}$. Larger $\ell_B$ corresponds to higher charge density of the PE. In the simulations, we use $a=1.12 \sigma$, $k=500.0\,\epsilon_0/\sigma^2$, $L=370\,\sigma$ and the temperature, $k_BT/\epsilon_0= 1$, is maintained through a  Nos\'{e}-Hoover thermostat. The long-ranged Coulomb interactions are evaluated using the particle-particle/particle-mesh (PPPM) technique, e.g.~\cite{anoop11,anoop12}. 

\begin{figure}
\includegraphics[width=\columnwidth]{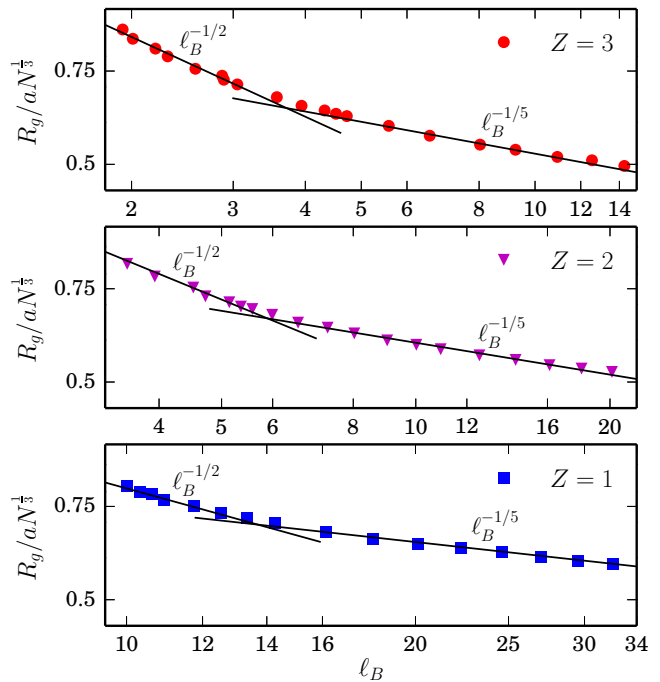}
\caption{Variation of the gyration radius $R_g$ of a PE chain with the reduced Bjerrum length $\ell_B$ for different valencies of  counterions. The chain length is $N=204$. The two power laws intersect at $R_g/aN^{1/3}\approx 0.63$\,($Z=3$), $0.66$ ($Z=2$) and $0.69$ ($Z=1$) with the corresponding crossover values $\ell_B^*(Z)\approx
3.71$ \,($Z=3$), $5.58$ ($Z=2$), and $13.70$ ($Z=1$).
}
\label{fig:RgZ}
\end{figure}

We now discuss the results from MD simulations of a single PE in a good solvent with purely
repulsive LJ interactions between all non-bonded pairs of monomers and counterions. The cutoff of the LJ interaction is set at $r_c=\sigma$, and the energy constant is $\epsilon=\epsilon_0$. We simulate the system for values of $\ell_B$ where the equilibrium configuration of a PE
is a collapsed state with $R_g \sim N^{1/3}$. The variation of the radius of gyration $R_g$ with $\ell_B$ in the globular regime is shown in Fig.~\ref{fig:RgZ}.
It can be seen from Fig.~\ref{fig:RgZ} that for $\ell_B <\ell_B^*(Z)$
the observed dependence,  $R_g \propto \ell_B^{-1/2} N^{1/3}$, is consistent with the predictions of the counterion-fluctuation theory~\cite{brilliantov98}. For $\ell_B > \ell_B^*(Z)$, we find a crossover to a different scaling, $R_g \propto \ell_B^{-1/5} N^{1/3}$, which is not predicted by any of the existing theories. The two regimes of  $\ell_B<\ell_B^*(Z)$ and $\ell_B>\ell_B^*(Z)$ will be referred to as weak and strong electrostatic regimes respectively.

\begin{figure}
\includegraphics[width=0.9\columnwidth]{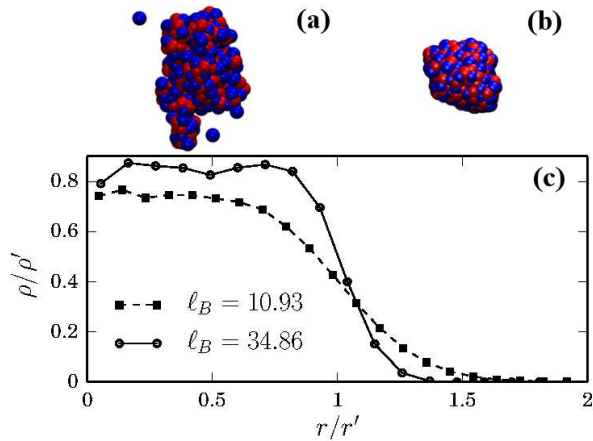}
\caption{Snapshots of collapsed PE in (a) weak electrostatic regime, $Z=1$, $\ell_B=10.93$  (left) and
(b) strong electrostatic regime, $Z=1$, $\ell_B=34.86$  (right).
(c) The corresponding radial number density profile $\rho$ of the counterions where $r$ is the distance of the counterion from the center of mass of the chain. $r^{\prime}$ is the distance at which the density is 50\% of the  density at $r=0$  and $\rho ^{\prime}=N/V^{\prime}$, where $V^ {\prime}=\frac{4}{3}\pi r^{\prime 3}$. }
\label{fig:WeakStrong}
\end{figure}

Typical snapshots of the system with monovalent counterions in weak and strong
electrostatic regimes are shown in Fig.~\ref{fig:WeakStrong}, which demonstrates that
the PE is much more compact in the strong electrostatic regime. Associated number density
profile of counterions measured from the centre of mass of the collapsed PE is also shown in
Fig.~\ref{fig:WeakStrong}. It can be seen that the profile has a broader tail in the weak electrostatic regime, suggesting that the counterions are more loosely bound.

We have verified that the exponents  and associated features seen in Fig.~\ref{fig:RgZ} are robust and independent on the details of the interaction by simulating two other good solvent conditions \cite{SuppInfo}:  (i) LJ interactions being  attractive ($r_c=2.5 \sigma$, $\epsilon=0.25 \,\epsilon_0$) for monomer-monomer pairs and purely repulsive
($r_c=1.0 \sigma$, $\epsilon=\epsilon_0$) for all other pairs and (ii) PE in the presence of explicit solvent molecules with attractive interactions between monomers and solvent ($r_c=2.5 \sigma$, $\epsilon=\epsilon_0$) and repulsive for all other pairs.
We also confirm that the results are independent of the length of the chain  for all values of $\ell_B$ that we have simulated (see the Supplementary Information).

The dependence of $R_g$ on $\ell_B$ in  the weak electrostatic regime supports the basic mechanism of the counterion-fluctuation theory as described in Ref.~\cite{brilliantov98}, where the PE free energy was truncated at the second virial coefficient. We now re-examine this theory to explain the dependence $R_g \propto \ell_B^{-1/5} N^{1/3}$ in the strong electrostatic regime by including more terms in the virial expansion
of the PE free energy: namely, we use the simplest generalisation of the counterion-fluctuation theory~\cite{brilliantov98}, including the third virial coefficient $C$.

\emph{Theory.}
The free energy of the system as a function of the radius of gyration $R_g$ of a PE chain can be written as~\cite{brilliantov98,khokhlov1994,BudkovJCP2015}
\begin{equation}
\label{eq:F}
F(R_g) =F_{\rm id. ch}(R_g)+F_{\rm vol}(R_g)+F_{\rm el}(R_g).
\end{equation}
Here $F_{\rm id.ch}(R_g)$ is the entropic part of the free energy corresponding
to the ideal chain~\cite{GrosbergKuznetsov_MM92,brilliantov98,khokhlov1994},
\begin{equation}
\label{eq:Fn}
\beta F_{\rm id.ch} \simeq  \frac94  \left(\alpha^2 + \alpha^{-2} \right),
\end{equation}
where $\alpha = R_g/R_{\rm g. id}$ is the expansion factor, with $R_{\rm g.id}$  being the radius of gyration of the ideal chain, $R^2_{\rm g.id}=Na^2/6$. $F_{\rm vol}$ refers to the volume interactions between the chain monomers, which may be written using the second and third virial coefficients as~\cite{khokhlov1994,BudkovJCP2015}:
\begin{equation}
\label{eq:vol}
\beta F_{\rm vol}=\left( \frac{N^2B}{2 V_{\rm g}} + \frac{N^3 C}{6 V_{\rm g}^2} \right)
 =\left( \frac{N^{1/2} \tilde{B}}{ \alpha^3 } + \frac{\tilde{C}}{\alpha^6 } \right)\, ,
\end{equation}
where $V_{\rm g} =(4\pi/3)R_g^3$ is the volume of gyration and we introduce the reduced virial coefficients, $\tilde{B} = 9 \sqrt{6} B /(4 \pi a^3)$ and $\tilde{C}= 81 C /(4 \pi^2 a^6)$. Finally, $F_{\rm el}$, which  takes into account all the electrostatic interactions (between the monomers and  counterions) as well as the entropic part of the counterions is given by~\cite{brilliantov98}:
\begin{eqnarray}
\label{eq:el}
\frac{\beta F_{\rm el}}{N}&=&\frac{3\sqrt{6} \ell_B N^{1/2}(1-\tilde{\rho})^2}{5\alpha} \left(1-\frac{2R_g}{3 R_0} \right) \\
&-&\frac{3}{Z}(1-\tilde{\rho}) \ln \left(\frac{R_0}{a}\right)
-\frac32 \left(\frac{2}{\pi^2} \right)^{1/3} \frac{\ell_B \sqrt{6} Z^{2/3} \tilde{\rho}^{4/3}}{N^{1/6} \alpha}. \nonumber
\end{eqnarray}
Here $\tilde{\rho}=\rho_{\rm in}/\rho_0$  with $\rho_{\rm in}$ being the number density of counterions within the volume occupied by the polymer chain $V_{\rm g}$ and $\rho_0=N_c/V_{\rm g}=N/(ZV_{\rm g})$ is the counterion density at the complete condensation. The value of $R_0$ quantifies the volume $4\pi R_0^3/3$ per chain in the solution and corresponds  to $L$ in  the MD simulations.
The above expression for $F_{\rm el}$ is valid for dilute solutions, $R_0 \gg R_g$ and for $N \gg 1$~\footnote{In Ref.~\cite{brilliantov98} the case of $Z=1$ has been addressed. A straightforward generalization yields Eq.~\eqref{eq:el}}. The first term in the right hand side of  Eq.~(\ref{eq:el}) accounts (on the mean-field level) for the electrostatic interactions in the system, while the second term describes the entropic part of the counterion free energy. The third term quantifies the contribution from electrostatic correlations to the free energy, and is absent within the Poisson-Boltzmann approximation~\cite{brilliantov98}.

We now focus on the globular state where $R_g\sim N^{1/3}$, so that $\alpha \ll 1$. In this case, the entropic part of the free energy $F_{\rm id.ch}$ [see Eq.~\eqref{eq:Fn}] may be ignored when compared to the other parts  of the free energy ($F_{\rm vol}+F_{\rm el}$).
Also, in the collapsed regime, most of the counterions are in the vicinity of the PE, which suggests the approximation $\tilde{\rho} \approx 1$ in Eq.~\eqref{eq:el}. Hence the electrostatic contribution to the free energy can be approximated as
\begin{equation}
\label{eq:elst}
\frac{\beta F_{\rm el}}{ N} \approx  -\frac{\tilde{Z}^2  \ell_B  }{N^{1/6} \alpha},
\end{equation}
where $\tilde{Z}^2 =(3/2)(2/\pi^2)^{1/3}\sqrt{6}\,Z^{2/3}$.
Thus for a single PE in any solvent, in the regime where the electrostatic contribution to the free energy dominates over the entropic one, Eqs.~(\ref{eq:vol}) and ~(\ref{eq:elst}) yield for the free energy:
\begin{equation}
\label{eq:ftot}
\frac{\beta F}{ N} = -\frac{\tilde{Z}^2  \ell_B  }{N^{1/6} \alpha}+\frac{\tilde{B}}{N^{1/2} \alpha^3} + \frac{\tilde{C}}{N\alpha^6}.
\end{equation}
Note that while Eq.~(\ref{eq:ftot}) takes into account the volume interactions between the chain monomers, such interactions with counterions may be also important for a dense globule. It is straightforward to take into account these interactions, which does not alter   the form of the free energy (\ref{eq:ftot}), but leads to the renormalization of $\tilde{B}$ and $\tilde{C}$ (see the Supplementary Information). For simplicity we keep the same notations for the renormalized coefficients.

In what follows we consider the case of a good solvent, which corresponds to positive coefficients $\tilde{B}$ and $\tilde{C}$. To find equilibrium $\alpha$ and hence $R_g$, one needs to minimize Eq.~(\ref{eq:ftot}) with respect to $\alpha$. The relative importance of the virial terms in Eq.~(\ref{eq:ftot}) depends on $N$, the virial coefficients $\tilde{B}$ and $\tilde{C}$, and the expansion factor $\alpha$.  The second virial term dominates when  $\alpha^3 > \tilde{C}N^{-1/2} /\tilde{B}$, which corresponds to the \emph{weak electrostatic regime}.
Neglecting the third virial term in Eq.~\eqref{eq:ftot} and minimizing $F$ with respect to $\alpha=R_g/R_{\rm g.\,id}$, we find
\begin{equation}
\label{eq:modcol}
R_g = \frac{\sqrt{\tilde{B}} a N^{1/3}}{\sqrt{2} \tilde{Z}\ell_B^{1/2}},
\end{equation}
as obtained in Ref.~\cite{brilliantov98}. This is consistent with the MD data for
$\ell_B<\ell_B^*$: $R_g\sim \ell_B^{-1/2}$, see Fig.~\ref{fig:RgZ}.

In contrast, in the \emph{strong electrostatic regime}, when $\alpha^3 < \tilde{C}N^{-1/2}/ \tilde{B}$, the third virial term is larger than the second one.  Hence, neglecting  the second virial term in Eq.~\eqref{eq:ftot}  and minimizing  the free energy, we obtain
\begin{equation}
\label{eq:strcol}
R_g = \frac{ \tilde{C}^{1/5} a N^{1/3}}{6^{3/10}\tilde{Z}^{2/5}  \ell_B^{1/5}}.
\end{equation}
This   scaling of $R_g$ is consistent with the MD simulation data for $\ell_B > \ell_B^*$: $R_g \sim \ell_B^{-1/5}$ as shown in  Fig.~\ref{fig:RgZ}.

\begin{figure}
\includegraphics[width=\columnwidth]{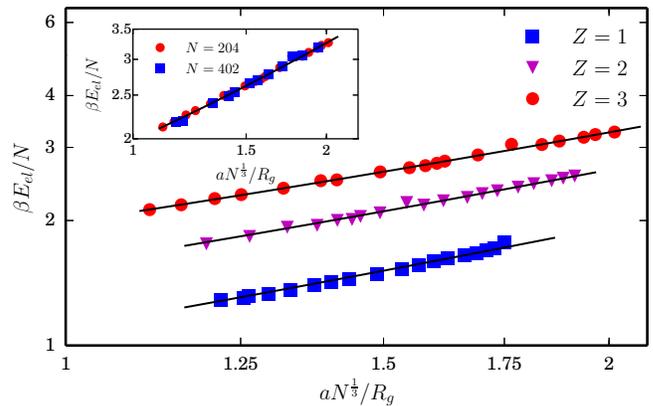}
\caption{Variation of the electrostatic energy $E_c$ with the radius of gyration $R_g$ for different valency. Inset:
Variation with the chain length $N$.  }
\label{fig:ElvsRg}
\end{figure}

To check independently our approximations for the electrostatic and the volume part of the free energy,  Eqs.~(\ref{eq:elst}) and (\ref{eq:vol}), we now calculate the respective components of the internal  energy and compare them to results from MD simulations. The electrostatic part of the internal energy $E_{\rm el} = \partial (\beta F_{\rm el} )/\partial \beta$ is given by
\begin{equation}
\label{eq:Eel}
\beta E_{\rm el}/( N \ell_B) = - \tilde{Z}^2 a N^{1/3}/\sqrt{6} R_g \sim  N^{1/3}R_g^{-1}.
\end{equation}
The scaling of $E_{el}$ as a function of $R_g$ is shown in Fig.~\ref{fig:ElvsRg} from the MD data, which clearly demonstrates the linear dependence of the electrostatic energy $E_{\rm el}$ on the inverse gyration radius $R_g$ as  obtained in Eq.~(\ref{eq:Eel}). We note that this linear dependence is valid in both weak and strong electrostatic regimes.

\begin{figure}
\includegraphics[width=\columnwidth]{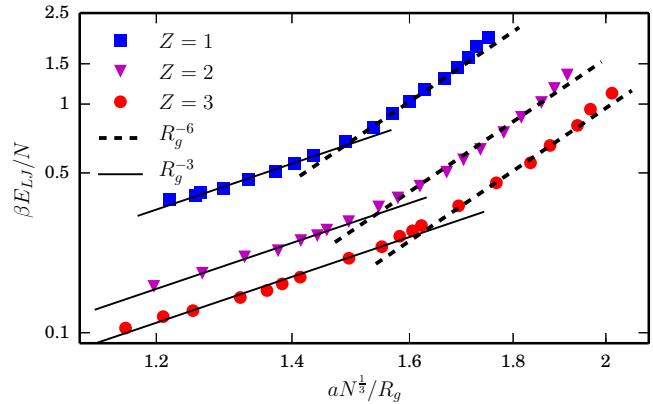}
\caption{Variation of the L-J energy of the system with  the radius of gyration $R_g$ for different valency.}
\label{fig:ELJ}
\end{figure}

Similarly, the internal energy corresponding to the volume interactions via LJ interactions, $E_{\rm LJ} = \partial (\beta F_{\rm vol} )/\partial \beta$, is given by
\begin{equation}
\label{eq:ELJ}
\beta E_{\rm LJ}/N = NB^{\prime} R_g^{-3} + N^2 C^{\prime}R_g^{-6},
\end{equation}
where $B^{\prime}=(3/8 \pi ) \beta \partial B /\partial \beta$ and    $C^{\prime}=(3/32 \pi^2) \beta \partial C /\partial \beta$. If the first term in the right hand side of Eq.~(\ref{eq:ELJ}) dominates, one obtains $E_{\rm LJ} \sim R_g^{-3}$; if the second one dominates, then $E_{\rm LJ} \sim R_g^{-6}$. In Fig.~\ref{fig:ELJ} we plot the respective internal energy due to volume interactions from our MD data. The figure convincingly illustrates the dominance of the second and third virial terms in the weak and strong electrostatic regimes correspondingly, with the
crossover occuring at $R_g/aN^{1/3} \approx 0.63$ ($Z=3$), $0.64$ ($Z=2$) and $0.68$ ($Z=1$). These values match closely with the crossover found in Fig.~\ref{fig:RgZ}.

\emph{Conclusion.} We elucidate the origin of attractive interactions
in a collapsed polyelectrolyte in a good solvent using MD simulations and theoretical analysis. We identify  two collapsed regimes, that we call as \emph{weak} and \emph{strong electrostatic regimes}. In the first regime the gyration radius $R_g$ of a chain scales with Bjerrum length $\ell_B$ as $R_g \sim \ell_B^{-1/2}$ while in the second one as $R_g \sim \ell_B^{-1/5}$.
This scaling is robust and independent on the valency of the counterions, volume interaction models between chain monomers and on the solvent models.
The observed scaling in the weak electrostatic regime ($R_g \sim N^{1/3} \ell_B^{-1/2}$) is not consistent with the predictions of the theories of fluctuating dipoles
($R_g \sim N^{1/3} \ell_B^{-2/3}$)~\cite{pincus98,Cherstvy2010,muthukumar2004,arti14}, or of the  amorphous ionic solid ($R_g \sim N^{1/3} \ell_B^{0}$)~\cite{delaCruz2000}, but agrees with the counterion-fluctuation theory~\cite{brilliantov98}.  At the same time the scaling in the strong electrostatic regime ($R_g \sim N^{1/3} \ell_B^{-1/5}$) is not consistent with any of the existing theories.

In this Letter, we modified the counterion-fluctuation theory~\cite{brilliantov98}, in which  density fluctuations of delocalised counterions  inside a chain globule  give rise to effective attractive interactions.  Including the third virial term into the volume-interaction part of the free energy of the chain $F_{\rm vol}$, we obtain the correct description for the $R_g (\ell_B)$ dependence in both weak and strong electrostatic regimes. We find that the different electrostatic regimes correspond to the dominance of different virial terms of $F_{\rm vol}$ and it may be envisaged that additional virial terms may be required at higher electrostatic strengths. We note that various theories explaining the origin of attractive interactions in a collapsed state of PE or PE gels \cite{brilliantov98,pincus98,Kardar1999,delaCruz2000,Cherstvy2010,muthukumar2004,arti14,Levin2002,Khokhlov1996,Kramarenko2000} differ mainly in the form of the electrostatic term. As we show in our MD simulations the scaling of the electrostatic energy with the gyration radius $R_g$ is the same for all values of $\ell_B$ and  is consistent with the counterion-fluctuation theory. Hence, our results strongly support the counterion-fluctuation mechanism of the PE collapse in a good solvent, suggested previously in Ref.~\cite{brilliantov98}.

\emph{Acknowledgments:} The simulations were carried out on the supercomputing machines Annapurna, Nandadevi and Satpura at the Institute of Mathematical Sciences.

\end{document}